\documentstyle[twocolumn,aps,epsf,floats]{revtex}

\begin{document}

\draft

\twocolumn[\hsize\textwidth\columnwidth\hsize\csname@twocolumnfalse\endcsname

\title {Low-field magnetoresistance in GaAs 2D holes}
\author{S. J. Papadakis\cite{SJPaddress}, E. P. De Poortere,
H. C. Manoharan\cite{HCMaddress}, J. B. Yau, M. Shayegan, and S.
A. Lyon}
\address{Department of Electrical Engineering, Princeton University,
Princeton, New Jersey  08544, USA.}
\date{\today}
\maketitle
\begin{abstract}
We report low-field magnetotransport data in two-dimensional hole
systems in GaAs/AlGaAs heterostructures and quantum wells, in a
large density range, $2.5 \times 10^{10} \leq p \leq 4.0 \times
10^{11}$ cm$^{-2}$, with primary focus on samples grown on (311)A
GaAs substrates. At high densities, $p \gtrsim 1 \times 10^{11}$
cm$^{-2}$, we observe a remarkably strong positive
magnetoresistance. It appears in samples with an anisotropic
in-plane mobility and predominantly along the low-mobility
direction, and is strongly dependent on the perpendicular electric
field and the resulting spin-orbit interaction induced
spin-subband population difference. A careful examination of the
data reveals that the magnetoresistance must result from a
combination of factors including the presence of two
spin-subbands, a corrugated quantum well interface which leads to
the mobility anisotropy, and possibly weak anti-localization. None
of these factors can alone account for the observed positive
magnetoresistance.  We also present the evolution of the data with
density:  the magnitude of the positive magnetoresistance
decreases with decreasing density until, at the lowest density
studied ($p = 2.5 \times 10^{10}$ cm$^{-2}$), it vanishes and is
replaced by a weak negative magnetoresistance.
\end{abstract}
\pacs{73}

\vskip1pc]

\section{Introduction}
\label{sec:intro}
\begin{figure}[tb]
\centerline{
\epsfxsize=3.0in
\epsfbox{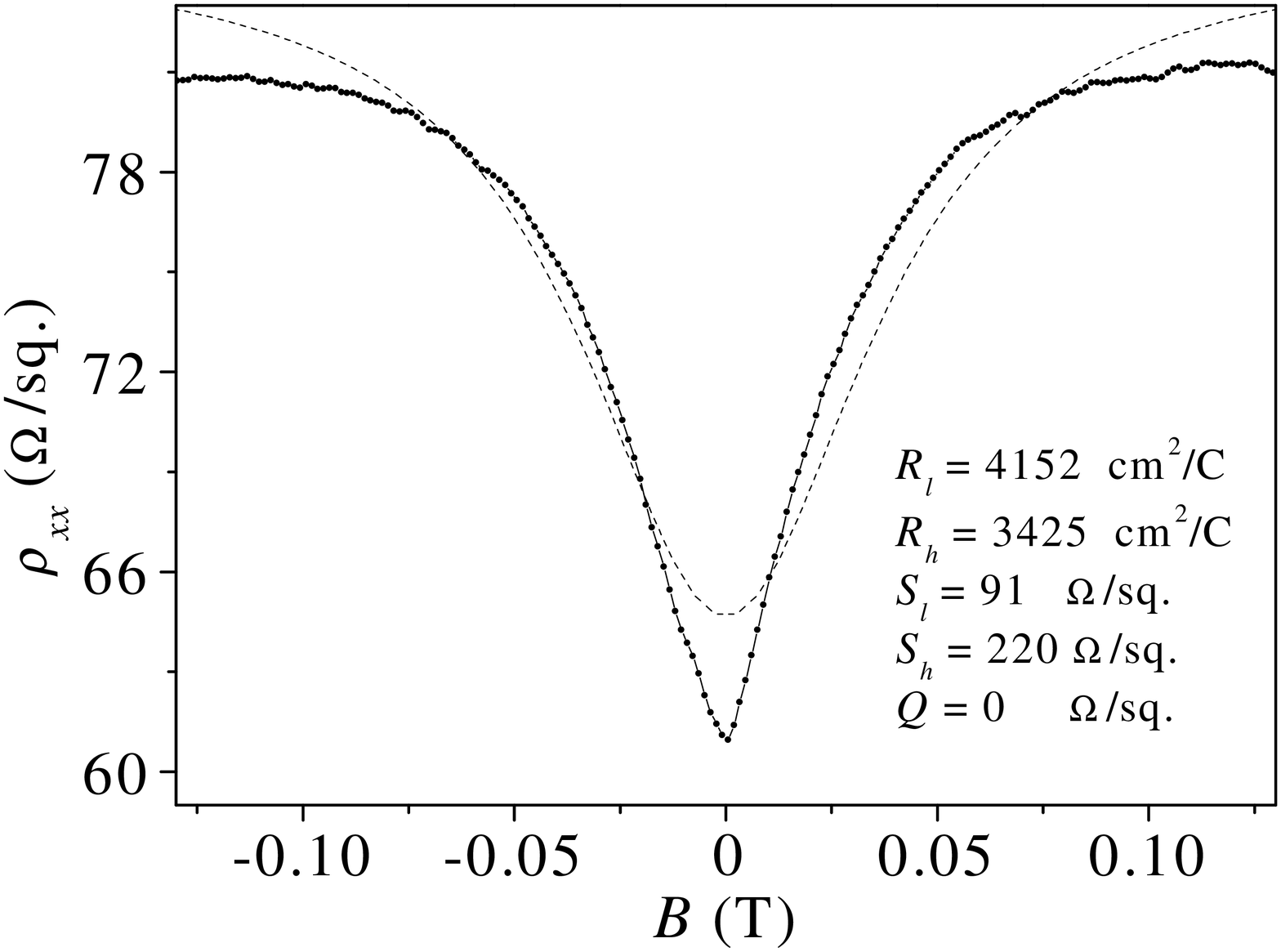}
}
\caption{Low-field magnetoresistance data at $T = 30$ mK from
sample A, a GaAs (311)A 2D hole system
with $p = 3.3 \times 10^{11}$ cm$^{-2}$, $E_{\perp} = -6$ kV/cm,
and current along $[01\bar1]$.  The dashed line is a fit of the
two-band model to the experimental data (see text), with the
parameters used for the fit given in the legend.}
\label{closeup}
\end{figure}

Two-dimensional (2D) hole systems confined to GaAs quantum wells
or GaAs/AlGaAs heterostructures often exhibit a strong positive
magnetoresistance (MR) at small perpendicular magnetic fields $|B|
\lesssim 0.1$ T
\cite{Eisenstein84,Murzin98,Papadakis99,Pedersen99,bothYaish}. An
example is shown in Fig. \ref{closeup}.
The origin of this MR is of renewed interest in light of the
puzzling metallic behavior recently observed in a number of 2D
systems including GaAs 2D holes
\cite{Murzin98,Papadakis99,bothYaish,Kravchenko94,Coleridge97,Lam97,Hanein98,Simmons98,Papadakis98}.
It has been argued that the positive MR can be explained by a
simple two-band model \cite{Murzin98,bothYaish}, where the two
bands arise from the spin-orbit induced spin-splitting of the
valence bands
\cite{Eisenstein84,Murzin98,Papadakis99,Pedersen99,bothYaish,spinsprev}.
Another report \cite{Pedersen99} relates the MR to weak
antilocalization which is expected to occur at low magnetic fields
in a system with spin-orbit coupling.  There is yet a third
possibility:  positive MR has also been observed in 2D carrier
systems with a one-dimensional periodic \cite{Weiss90,Beton90}, or
nearly periodic \cite{Akabori00}, potential modulation. This
suggests that the unintentional corrugations which are often
present at GaAs/AlGaAs surfaces and interfaces may also contribute
to the observed MR.

In this paper we report extensive data on 2D holes in GaAs quantum
wells and GaAs/AlGaAs heterostructures, grown on (311)A and (100)
GaAs substrates, as well as a 2D {\em electron} system grown on a
(311)A GaAs substrate.  The (311)A 2D hole data, which are the
main focus of this paper, span a density ($p$) range of $2.5
\times 10^{10} \leq p \leq 4.0 \times 10^{11}$ cm$^{-2}$.  The
results collectively reveal the following trends. At high
densities and in the presence of a large electric field
perpendicular to the plane of the (311)A 2D hole system, we
observe a pronounced positive MR when the current is driven along
the $[01\bar1]$ direction.  A much weaker MR is observed for
current along the $[\bar233]$.  As the perpendicular electric
field is reduced, the MR diminishes but does not vanish.  On the
other hand, as the 2D density is lowered the positive MR
progressively becomes smaller and vanishes at the lowest measured
density.   The behavior of the MR for current along $[01\bar1]$
cannot be explained by any one of the previously mentioned three
mechanisms. The combination of all three, however, is likely to
account for the MR.  We also measure the temperature dependence of
the resistivity at $B = 0$ and discuss its relation to the
observed MR.

The paper is organized as follows.  Section \ref{sec:expt}
describes the experimental setup and the samples used in our
study. Section \ref{sec:high-dens} concentrates on the
high-density data, and demonstrates how each of the
above-mentioned mechanisms for positive MR is expressed in the
data.  Section \ref{sec:low-dens} shows the evolution of the MR as
the density is reduced.  In Section \ref{sec:Tdeps} we make a
comparison between the MR and the $B = 0$ temperature dependence
of the resistivity where metallic behavior is observed.  Section
\ref{sec:Summ} summarizes our results and conclusions.

\section{Experimental details}
\label{sec:expt}

We studied the MR in square quantum wells (QWs) as well as in
single heterojunction samples, grown by molecular beam epitaxy
(MBE) on both (311)A and (100) insulating GaAs substrates.  We
report here data for samples from six different wafers; a summary
of the important sample parameters is given in Table
\ref{samples}.
\begin{table*}
\caption{Summary of sample parameters.  Densities and mobilities are for ungated samples.}
\label{samples}
\begin{tabular}{|c|c|c|c|c|c|c|c|}
Sample & Substrate & Carrier & Dopant & density ($10^{11}$
cm$^{-2}$) & $\mu_{xx}$ ($10^5$ cm$^2$/Vs) & $\mu_{yy}$ ($10^5$
cm$^2$/Vs) & QW structure \\ \hline
A & (311)A & holes           & Si & 2.2 & 4.0 & 5.8 & square     \\ \hline  
B & (311)A & holes           & Si & 2.2 & 5.5 & 8.0 & square     \\ \hline  
C & (100)  & holes           & Be & 2.0 & 1.5 & 3.7 & square     \\ \hline  
D & (311)A & electrons       & Si & 1.5 & 2.9 & 3.2 & triangular \\ \hline  
E & (311)A & holes           & Si & 1.5 & 0.7 & 4.2 & triangular \\ \hline  
F & (311)A & holes           & Si & 0.9 & 3.3 & 4.9 & square     \\         
\end{tabular}
\end{table*}
The samples are modulation-doped with either Si or Be.  Use of Be
as a $p$-type (acceptor) dopant for growth on GaAs (100)
substrates is standard, although the quality of the 2D hole system
may be somewhat compromised because of the fast Be diffusion in
GaAs, and a memory effect in the MBE chamber.

Alternatively, one can grow the heterostructure on GaAs (311)A
substrates and use Si which, under normal growth conditions, is
incorporated as an acceptor on this surface.  Such samples are
known to show exceptionally high quality as measured, e.g., by
their low-temperature mobility \cite{SantosNote}. They comprise
the majority of the samples used in our work as well as that of
others who have studied the metallic behavior in GaAs 2D hole
systems \cite{Papadakis99,Hanein98,Simmons98}.  The GaAs/AlGaAs
(311)A interface, however, has quasi-periodic corrugations along
the $[\bar233]$ direction \cite{Wassermeier95}, and the
corrugations lead to an in-plane mobility anisotropy
\cite{Heremans94}: mobility is typically larger along the
$[\bar233]$ direction by a factor of 2 or 3 compared to the
mobility along $[01\bar1]$.  This mobility anisotropy is likely
related to an anisotropic interface roughness scattering
\cite{Heremans94} and as we will show here, plays an important
role in the low-field MR.

For comparison, we also studied a modulation-doped (311)A
GaAs/AlGaAs heterojunction sample containing 2D {\it electrons}.
This sample was grown under MBE growth conditions similar to those
used for the 2D hole systems:  the substrate temperature during
the growth of the GaAs/AlGaAs interface and AlGaAs spacer was kept
the same as in 2D hole samples, and was reduced only when the
dopant (Si) atoms were introduced so that they were incorporated
as donors \cite{Agawa94}. The interface morphology of the 2D
electron system in this sample should therefore closely resemble
that of the 2D hole samples.

The samples were patterned with $L$-shaped Hall bars allowing
simultaneous measurements of longitudinal ($\rho_{xx}$ and
$\rho_{yy}$) and transverse ($\rho_{xy}$) magnetoresistances along
two different current directions.  On the (311)A samples, the arms
of the Hall bar were aligned along [$01\bar1$] and [$\bar233$],
which are respectively the low- and high-mobility ($\mu$)
directions. For purposes of discussion in this paper, we will
refer to the resistivity of the low-$\mu$ direction as $\rho_{xx}$
and the resistivity of the high-$\mu$ direction as $\rho_{yy}$. On
the (100) samples the arms of the Hall bar were aligned along
[011] and [$01\bar1$].  Measurements were done in dilution and
$^3$He refrigerators with base temperatures ($T$) of 30 and 300 mK
respectively. We measured the resistivity with a conventional
low-frequency lock-in technique using currents of 1-10 nA. Samples
A, B, and F had metallic front and back gates so that we could
independently control both the 2D hole density and the electric
field applied perpendicular to the QW ($E_{\perp}$)
\cite{Papadakis99}.  $E_{\perp}$ is defined as positive when the
electric field is pointing from the substrate towards the front
gate.

\section{Magnetoresistance at high density}
\label{sec:high-dens}

\begin{figure}[tb]
\centerline{
\epsfxsize=3.25in
\epsfbox{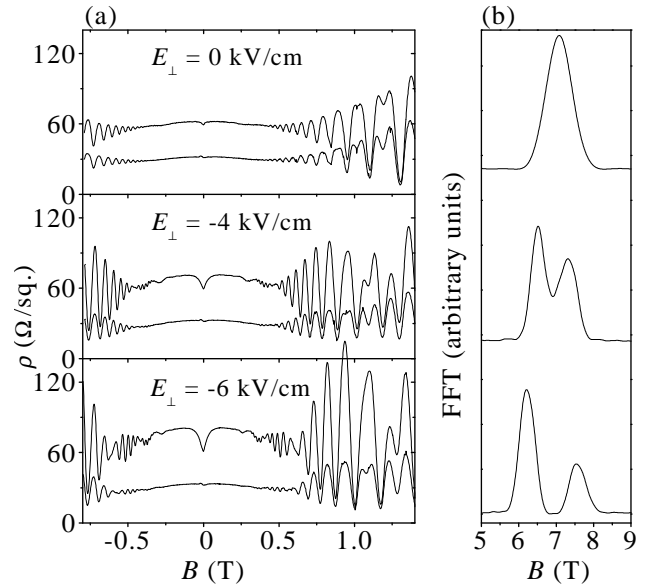}
} \vskip-6.5pc
\caption{a)  Magnetoresistance data at $T = 30$ mK from sample A with
$p = 3.3 \times 10^{11}$ cm$^{-2}$, shown for various $E_{\perp}$.
The low-field magnetoresistance is strong when the current is
along the low-$\mu$ $([01\bar1])$ direction (upper trace in each
panel), while it is much weaker when the current is along the
high-$\mu$ $([\bar233])$ direction (lower trace).  b) Fourier
transforms of the Shubnikov-de Haas oscillations.}
\label{M340skew}
\end{figure}

\begin{figure}[tb]
\centerline{
\epsfxsize=2.8in
\epsfbox{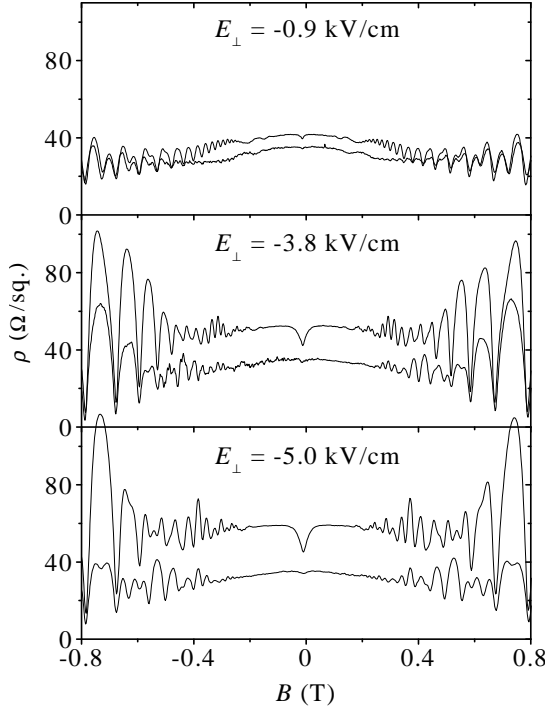}
}\vskip0.3pc
\caption{Magnetoresistance data at $T = 30$ mK from sample B, a 2D hole
system in a square QW grown on a (311)A substrate with $p = 2.3
\times 10^{11}$ cm$^{-2}$, shown for various $E_{\perp}$.  The
upper trace in each panel is from the low-$\mu$ ($[01\bar1]$)
direction, while the lower trace is from the high-$\mu$
($[\bar233]$) direction.}
\label{M289skew}
\end{figure}

The positive MR we observe in these GaAs 2D hole samples has
several remarkable properties. The most striking is that it can be
quite large.  In Fig. \ref{closeup}, the longitudinal resistivity
$\rho_{xx}$ rises by 33\% from $B = 0$ to $|B| = 0.1$ T.
$\rho_{xx}$ also often shows a cusp, which is sharp to our
measurement resolution, at $B = 0$. Furthermore, we have observed
the positive MR in samples that display a mobility anisotropy, and
then often only, and always much more strongly, in the
low-mobility direction (Figs. \ref{M340skew} and \ref{M289skew}).
We will focus initially on data from 2D hole systems in square QWs
grown on (311)A samples, and then present and compare data from
the other systems.

Our data reveal that the magnitude of the MR is influenced by the
symmetry of the potential that confines the 2D holes. Figures
\ref{M340skew} and \ref{M289skew}
show MR data for samples A and B, from two different wafers, as
the asymmetry of the QW potential is changed. The mobilities of
these samples are typically in excess of 500,000 cm$^2$/Vs, and
the high quality is evident from the strong Shubnikov-de Haas
oscillations. When the QW is made asymmetric by the application of
a perpendicular electric field $E_{\perp}$, the $B = 0$
spin-splitting grows.  The method of tuning $E_{\perp}$ is
described in detail in Ref. \onlinecite{Papadakis99}. Beating, due
to the presence of two spin-subbands with significantly different
populations, is evident in the traces at larger $|E_{\perp}|$
\cite{Papadakis99,spinsprev}. Examples of Fourier transforms of
$\rho_{xx}$ vs. $B^{-1}$ are shown in Fig. \ref{M340skew}b: the
single peak at low $|E_{\perp}|$ splits into two peaks at larger
$|E_{\perp}|$, giving a quantitative measure of the
spin-splitting.  The magnitude of the MR feature is correlated
with the spin-splitting. Here we show data only for $E_{\perp} <
0$, which corresponds to an electric field pointing towards the
substrate. Data with $E_{\perp} > 0$ show the same behavior
\cite{Papadakis99}: both the spin-subband densities and the
magnitude of the MR are symmetric around $E_{\perp} = 0$.

\begin{figure}[tb]
\centerline{
\epsfxsize=2.5in
\epsfbox{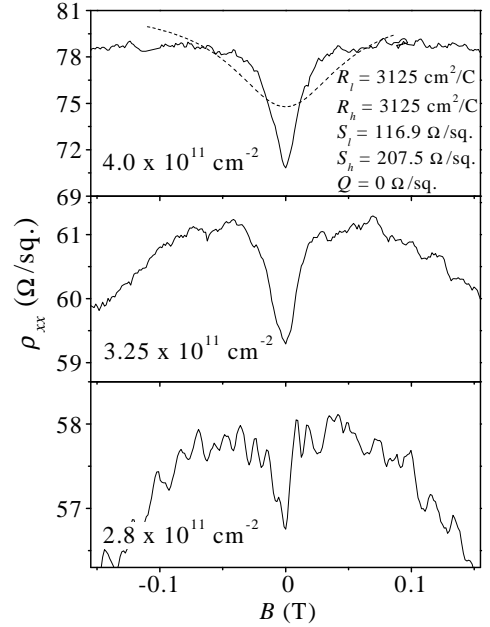}
}
\caption{Magnetoresistance at $T = 30$ mK along the low-$\mu$
direction ($\rho_{xx}$) from sample A as density is reduced;
$E_{\perp} \simeq 0$ for all three traces. Results of the two-band
model fit are shown in the top panel (dotted curve).}
\label{M340dens}
\end{figure}

As demonstrated in Fig. \ref{M340dens}, the magnitude of the MR is
affected by changes in the 2D hole density as well.  A reduction
in the density, with $E_{\perp}$ held constant, causes a reduction
in the magnitude of the MR. There is very little change in its
shape (Fig. \ref{M340dens}); and it retains its cusp at $B = 0$
until it is almost indistinguishable from the noise.
The behavior of the MR over a wider density and $E_{\perp}$ range
is discussed in Section \ref{sec:low-dens}.

\begin{figure}[tb]
\centerline{
\epsfxsize=2.5in
\epsfbox{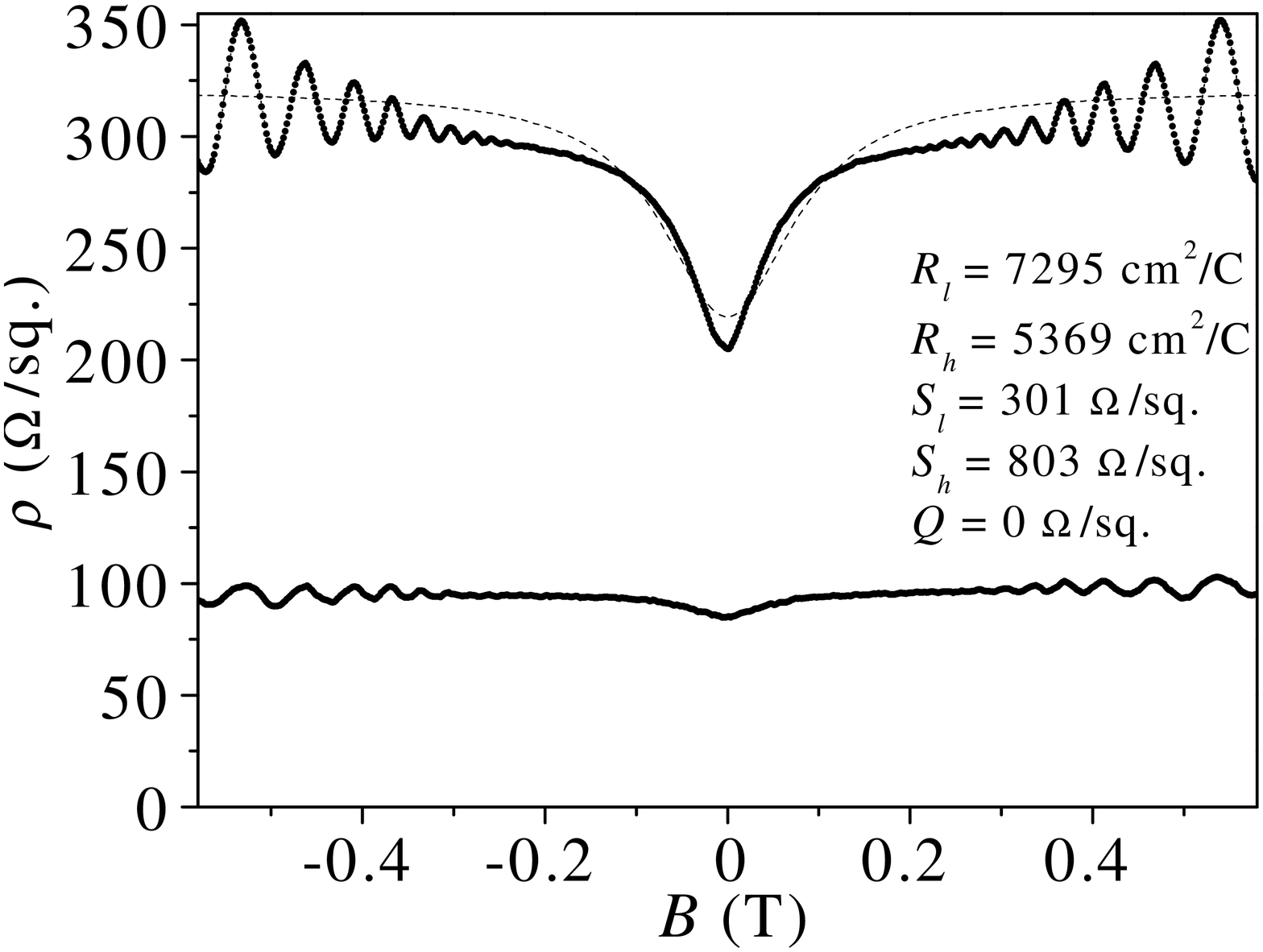}
}
\caption{Low field magnetoresistance data at $T = 300$ mK from sample C, a 2D hole system
grown on a (100) substrate with $p = 2.0 \times 10^{11}$
cm$^{-2}$.  The results of the two-band model fit for the top
trace are also shown (dotted curve).}
\label{M371closeup}
\end{figure}

The MR is also observed in Be-doped samples grown on GaAs (100)
substrates. An example is shown in Fig. \ref{M371closeup}.
Here, positive MR is observed in both $\rho_{xx}$ and $\rho_{yy}$,
but it is again much smaller in $\rho_{yy}$.  In the (100)
samples, the MR feature depends on $E_{\perp}$ and density in the
same way it does in the (311)A samples.  The strong similarities
between the positive MR observed in (100) and (311)A samples
suggest that the feature has a similar origin in both systems.

Here we discuss three possible origins for the low-$B$ positive
MR.  Weak anti-localization \cite{Pedersen99}, a two-band model
\cite{Murzin98,bothYaish}, and a one-dimensional periodic
potential modulation \cite{Akabori00} have all been invoked to
explain positive MR in the past.  Our comprehensive set of data
shows that some combination of these is necessary to describe the
MR in GaAs 2D holes.

A much weaker MR with a similar shape has been seen in 2D holes on
(100) substrates and attributed to weak anti-localization
\cite{Pedersen99}.  The cusp at $B = 0$ in our data is suggestive
of weak anti-localization, but the magnitude of the resitivity
change is many orders of magnitude too large to be caused by weak
anti-localization.  Weak localization and anti-localization
typically contribute a correction of order $\sim 0.01e^2/h$ to the
conductivity of a 2D system.  The conductivities of our samples
are orders of magnitude larger than $e^2/h$, so such a correction
is far smaller than the total change in the conductance we
observe.

\begin{figure}[tb]
\centerline{
\epsfxsize=2.5in
\epsfbox{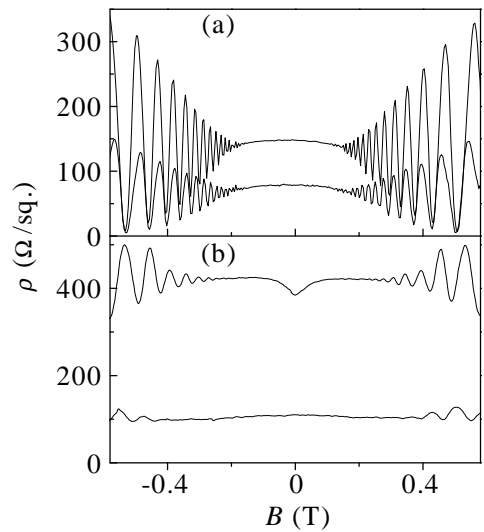}
}
\caption{a) Low-$B$ magnetoresistance data at $T = 30$ mK from both the low- and high-$\mu$
directions of sample D, a (311)A 2D {\it electron} system in a
triangular well with a density of $1.5 \times 10^{11}$ cm$^{-2}$.
There is a mobility anisotropy, but no low-$B$ magnetoresistance
feature. b) Low-$B$ magnetoresistance data at $T = 300$ mK from
sample E, a (311)A 2D hole system in a triangular well with $p =
1.68 \times 10^{11}$ cm$^{-2}$. This sample shows a strong
magnetoresistance feature in the low-mobility ($[01\bar1]$)
direction.}
\label{nodimple}
\end{figure}

In GaAs 2D holes on (100) substrates, Murzin {\it et al.}
\cite{Murzin98} and Yaish {\it et al.} \cite{bothYaish} have
observed a similar MR feature and described it using a simple
two-band model.  This two-band model initially looks promising in
our system as well.  We have already seen that the magnitude of
the positive MR increases when the spin-subband population
difference is increased.  Another factor that qualitatively
supports the two-band hypothesis is that the MR is completely
absent in a 2D electron system in a triangular well grown on a
(311)A substrate; 2D electrons in GaAs have a very small
spin-orbit coupling so they occupy essentially a single
spin-degenerate subband.  Figure \ref{nodimple}a shows data from a
2D electron system on (311)A, and Fig. \ref{nodimple}b shows data
from a similar (311)A 2D hole system for comparison.

We have fit the two-band model as used by Yaish {\it et al.}
\cite{bothYaish} to our data. Examples are shown in Fig.
\ref{closeup}, the top panel of Fig. \ref{M340dens}, and in Fig.
\ref{M371closeup}.  We multiply the Shubnikov-de Haas frequencies
deduced from the Fourier transforms of $\rho_{xx}$ vs. $B^{-1}$ by
$h/e^2$ to get the two subband densities (when there is one
measured frequency, we use the same density for both subbands),
from which we calculate the Hall coefficients $R_{l,h} =
1/p_{l,h}e$ for the light ($l$) and heavy ($h$) holes
\cite{SdHfreq}. These Hall coefficients are used as inputs for the
two-band model.  From the fits we extract $S_l$, $S_h$, and $Q$,
which can be expressed using elements of the conductivity tensor:
$S_l = \sigma_{ll}^{-1} + \sigma_{lh}^{-1}$, $S_h =
\sigma_{hh}^{-1} + \sigma_{hl}^{-1}$, and $Q =
\lambda\sigma_{lh}^{-1} = \lambda^{-1}\sigma_{hl}^{-1}$. The
$\sigma_{ii}$ take into account intraband scattering and the
$\sigma_{ij}$ account for interband scattering.

We find that, overall, the model produces a less satisfactory fit
to our data than to that of Yaish {\it et al.}.  The values it
produces for $S_l$ and $S_h$ are typically within a factor of two
of each other, which are reasonable for two spin-subbands.
However, the best fit to our data is always for $Q = 0$, which
corresponds to no intersubband scattering, and the model always
produces a wider MR feature than the data.  We have constrained
$Q$ such that $Q \ge 0$, since the conductivity due to
intersubband scattering rate cannot be negative.  Intersubband
scattering can only increase the width of the feature, so the fact
that our data shows sharper MR than the two-band model with no
intersubband scattering implies that this simple model is not
adequately explaining the MR. Furthermore, the two-band model
cannot match the cusp seen in the data at $B = 0$.

Positive MR has also been observed repeatedly when a
one-dimensional periodic potential is applied to a 2D system (see
\onlinecite{Akabori00} and references therein.).  The qualitative
characteristics of our data are similar to those of these systems.
In our case, we have not imposed an intentional periodic
potential, but the resistance anisotropy of our system is
suggestive of some sort of anisotropic disorder.  Transport data
in (311)A samples indeed suggest that interface-roughness
scattering limits the mobility along the $[01\bar1]$ direction
\cite{Heremans94}, and scanning-tunneling-microscopy (STM) images
show one-dimensional corrugations along the $[\bar233]$ direction
\cite{Wassermeier95}. These images show two sets of corrugations:
one nearly regular with a 32 {\AA} period and a 2 {\AA} amplitude,
and one quasi-periodic with a $\sim 1000$ {\AA} period and up to
$\sim 15$ {\AA} amplitude.  For samples grown on (100) substrates,
it is initially a surprise that there is a mobility anisotropy at
all, since the [011] and $[01\bar1]$ directions are expected to be
crystallographically equivalent. However, any slight miscut of the
substrate from the ideal (100) direction could break the symmetry,
and indeed STM images have shown a one-dimensional quasi-periodic
potential, with a period of a few hundred {\AA}, to exist in
samples grown on epi-ready GaAs (100) wafers as well
\cite{Wassermeier96}.  Also, AFM studies of high-quality (100)
samples have shown an anisotropic surface roughness that
correlates with anisotropy in transport in the 2D layer
\cite{Willett01}. We have not independently verified that in our
(100) samples the low-mobility direction is indeed perpendicular
to the corrugations, but it is a reasonable assumption given the
resemblance to the (311)A data.

The similarity between our data and the data of Akabori {\it et
al.} from a corrugated 2D electron system in InGaAs is remarkable
\cite{Akabori00}. Most obviously, in both systems the current
direction perpendicular to the corrugations has a lower mobility.
Also, in both systems, the positive MR is strong for current
perpendicular to the corrugations. Furthermore, the amplitude of
the Shubnikov-de Haas oscillations is larger for current
perpendicular to the corrugations than for current parallel (see
our Figs. \ref{M340skew} and \ref{M371closeup}, and Fig. 3a of
Akabori {\it et al.}).  Finally, the fact that the MR grows when
$|E_{\perp}|$ is made larger is consistent with the hypothesis
that the periodic modulation of the interface plays a role. When
$E_{\perp} = 0$, the carrier wavefunction is centered in the QW,
so it has little overlap with the barriers at the edges of the QW.
When $|E_{\perp}|$ is increased, the wavefunction is pushed
towards one of the barriers, so any interface corrugations present
will have a larger effect.

We now provide a quantitative assessment of this hypothesis.
Akabori {\it et al.} apply the semiclassical model of Beton {\it
et al.} \cite{Beton90}, which relates the period and amplitude of
the potential corrugation to the $B$ at which the positive MR
saturates. Adopting such a model and assuming corrugation periods
of 32 {\AA} and 1000 {\AA} in our (311)A samples
\cite{Wassermeier95}, we calculate corrugation amplitudes of 0.002
and 0.07 meV respectively.  These values are much smaller than the
typical Fermi energy $E_F$ of 2 meV for 2D holes at a density of
$3.3 \times 10^{11}$ cm$^{-2}$. This model therefore gives an
inconsistent picture, since potential modulations more than an
order of magnitude smaller than $E_F$ would not be expected to
produce such a large MR.  A quantum mechanical model
\cite{Weiss90,Streda90}, on the other hand, yields potential
amplitudes of 0.6 and 0.1 meV respectively. These values are
reasonable, and can possibly lead to the observed MR.

The comparisons between the data of Akabori {\it et al.} and ours
suggest that the interface corrugations in our system are playing
a role in the positive MR, but once again, looking at the complete
set of data shows that the corrugations alone cannot be the origin
of the positive MR in our samples. As mentioned above, data from
the 2D {\it electrons} on (311)A substrates show no positive MR
(Fig. \ref{nodimple}a). The growth of the QW in these samples is
identical to the growth of the QW in (311)A 2D hole systems, so we
expect the interfaces to be the same as well. The only difference
in the samples is that, after the QW has already been grown, the
substrate temperature is reduced while the Si dopant atoms are
being deposited, causing Si to become a donor instead of an
acceptor. An important difference between 2D electrons and 2D
holes in GaAs is that spin-orbit interaction is strong for the 2D
holes, but weak for the 2D electrons.  Therefore, it appears that
the existence of both interface corrugations and two spin-subbands
is necessary for the expression of the positive MR we have
observed. We note that in the system of Akabori {\it et al.}, 2D
electrons in InGaAs, there is also a strong spin-orbit
interaction.

In short, the data we have shown collectively reveal that a
combination of factors is responsible for the low-$B$ positive MR
we observe.  It appears to be due to the presence of {\it both}
two occupied subbands {\em and} a one-dimensional quasi-periodic
potential modulation. Furthermore, neither of these mechanisms is
likely to explain the sharp cusp we observed at $B = 0$.  It is
possible that weak anti-localization may also be playing a role in
the very-low-$B$ regime, causing this cusp.

\section{Magnetoresistance at low density}
\label{sec:low-dens}

\begin{figure*}[tb]
\centerline{
\epsfxsize=6.9in
\epsfbox{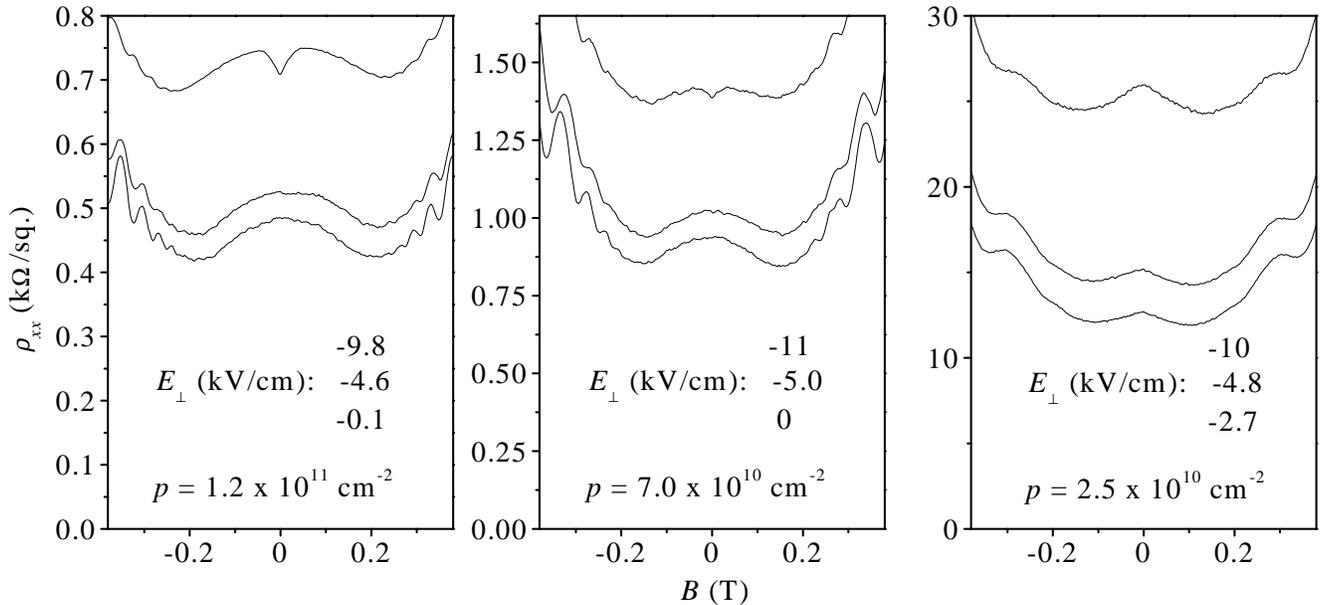}
}
\caption{Low-$\mu$ ($\rho_{xx}$) traces at $T = 30$ mK from sample F
at three $E_{\perp}$ for each of three densities.}
\label{low-dens}
\end{figure*}

In this section we show how the low-$B$ MR of 2D holes evolves as
the density is reduced.  Figure \ref{low-dens} shows $\rho_{xx}$
traces from sample F for three different $E_{\perp}$ at each of
three densities.
Raw data are plotted, without any shifts.  The curves at larger
$|E_{\perp}|$ have lower mobility because as $|E_{\perp}|$ is
increased, the wavefunction is pushed to one side of the QW, where
in this current direction it is more strongly affected by the
interface corrugations.  As mentioned above, this could be a
factor in the increasing strength of the positive MR as well.

At the highest density $p = 1.2 \times 10^{11}$ cm$^{-2}$, a
positive MR is clearly visible only for the largest $|E_{\perp}|$,
$E_{\perp} = -9.8$ kV/cm.  At lower $|E_{\perp}|$, it is not
visible: at low $B$ there is only a weak negative MR.  At the
intermediate $p$ of $7.0 \times 10^{10}$ cm$^{-2}$, the MR feature
is still visible at the highest $|E_{\perp}|$, but is noticeably
smaller.  When $p = 2.5 \times 10^{10}$ cm$^{-2}$, there is no
positive MR even at the largest $|E_{\perp}|$ of -10 kV/cm.  In
this range, there is only a {\em negative} MR, which in a similar
system has been attributed to weak localization \cite{Simmons99}.
Figure \ref{low-dens} data show that for asymmetric QWs, the
positive MR feature can be seen to quite low densities, well below
the density at which two frequencies can no longer be resolved in
the Shubnikov-de Haas oscillations.

\section{Temperature dependence of the $B = 0$ resistivity}
\label{sec:Tdeps}

\begin{figure}[tb]
\centerline{
\epsfxsize=3.1in
\epsfbox{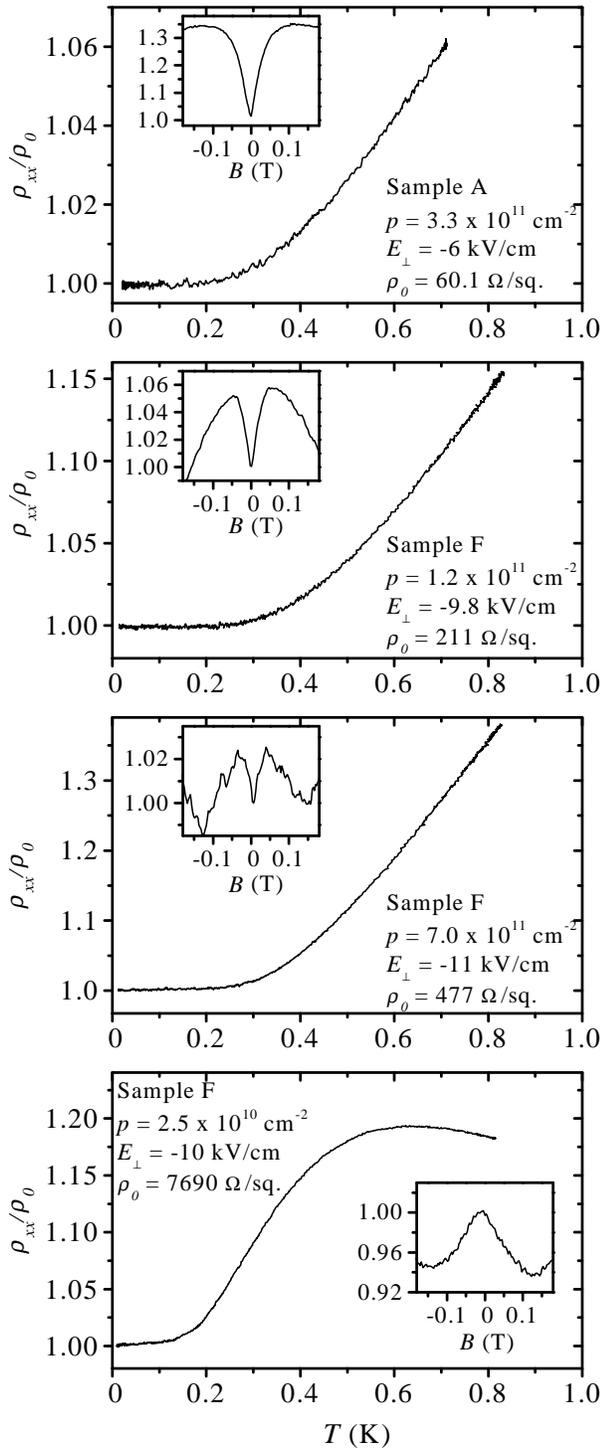}
} \vskip0.3pc
\caption{Temperature dependence of $\rho_{xx}$ at $B = 0$ for
four densities from samples A and F.  The insets show low-$B$ MR
traces at $T = 30$ mK.  The $y$-axes are all plotted as
$\rho_{xx}/\rho_0$, where $\rho_0$ is $\rho_{xx}$ at $B = 0$ and
$T = 30$ mK.}
\label{Tdeps}
\end{figure}

It is interesting to explore the relation between the MR and the
temperature dependence of the $B = 0$ resistivity which has been
controversial recently
\cite{Murzin98,bothYaish,Kravchenko94,Coleridge97,Lam97,Hanein98,Simmons98,Papadakis98}.
In Fig. \ref{Tdeps}, we show the metallic $T$-dependence of
$\rho_{xx}$ at $B = 0$ and the MR feature for comparison.  Both
the MR and the $B = 0$ $T$-dependence of $\rho_{xx}$ are plotted
as a fractional change $\rho_{xx}/\rho_0$ (where $\rho_0$ is
$\rho_{xx}$ at $B = 0$ and $T = 30$ mK) so that the relative
magnitudes can be easily compared.
It has been proposed that the $B = 0$ $T$-dependence of
$\rho_{xx}$ can be explained in the two-band model
\cite{Murzin98,bothYaish}. It is only at high $p$, where the rise
in $\rho$ as $T$ is increased is smaller than the MR, that our
data are consistent with this hypothesis.  For $p = 3.3 \times
10^{11}$ cm$^{-2}$, $\rho_{xx}$ shows a 6\% rise as $T$ increases
from 30 mK to 800 mK, while the positive MR from $B = 0$ to 0.1 T
is 33\%. As $p$ is reduced however, the two features show opposite
trends, and the data do not support the two-band hypothesis. For
smaller $p$, the change in $\rho_{xx}$ with increasing $T$ becomes
larger while the MR becomes smaller. For $p = 1.2 \times 10^{11}$
cm$^{-2}$, $\rho_{xx}/\rho_0$ at $T = 0.8$ K is 3 times larger
than the positive MR.  At $p = 7.0 \times 10^{10}$ cm$^{-2}$, the
MR is only 2\%, and $\rho_{xx}/\rho_0$ at $T = 0.8$ K has risen to
nearly 40 \%.  At the lowest density, no positive MR is observed,
yet there is still strong metallic behavior in the $T$-dependence
of $\rho_{xx}$ at $B = 0$.  There is evidence that the existence
of two spin-subbands does play a role in the $B = 0$ metallic
behavior \cite{Papadakis00b,Papadakis00d} throughout the density
range in which it is observed, but a comparison of the low-$B$ MR
feature with the $B = 0$ $T$-dependence of $\rho$ suggests that a
simple two-band model with intersubband scattering is not
sufficient to explain the relationship, as the metallic behavior
is strongest at densities where the MR is very weak or altogether
absent.

\section{Summary}
\label{sec:Summ}

In summary, we observe a positive MR in high-density 2D hole
systems in GaAs quantum wells.  The MR appears to be related to
both the spin-splitting in the system {\it and} to quasi-periodic
corrugations perpendicular to the current direction at the
interfaces between the quantum wells and the barriers.  The data
also show a cusp at $B = 0$ that is suggestive of weak
anti-localization.  At a fixed perpendicular electric field, the
MR is reduced in size when the 2D hole density is reduced.
Finally, the magnitude of the MR is not correlated with the
magnitude of the rise in sample resistivity with increasing $T$ at
$B = 0$.  This implies that the metallic behavior is not driven
solely by the factors that lead to the MR.

We would like to thank J. P. Eisenstein and B. L. Altshuler for
fruitful discussions.  This project was supported by the NSF, the
ARO, and the DOE.


\end{document}